\documentclass[aps,nofootinbib,onecolumn,citeautoscript]{revtex4}
\usepackage{amsmath,amssymb}
\usepackage[colorlinks=true,linkcolor=red,citecolor=blue]{hyperref}
\usepackage{breqn}
\usepackage{enumerate}
\usepackage{graphicx}
\usepackage{float}
\graphicspath{{/home/kapil/My_data/kapilpaper/arXiv/spin_squeezing/}}
\DeclareGraphicsExtensions{.pdf,.eps}

\date{\today}
\begin{document}
\title{\Large Decoherence induced spin squeezing  signatures in Greenberger\textendash Horne\textendash Zeilinger and W states}

\author{Kapil K. Sharma$^\ast$  \\
\textit{$^\ast$ Department of Electrical Engineering, \\Indian Institute of Technology Bombay, \\ Mumbai 400076, India} \\
E-mail: $^\ast$iitbkapil@gmail.com\\ 
%\textit{$^\dagger$address of xyz, \\ Allahabad-211004, India.} \\
%Email: $^\dagger$mail of xyz
}

\begin{abstract}
We reckon the behaviour of spin squeezing in tripartite unsqueezed maximally entangled Greenberger\textendash Horne\textendash Zeilinger (GHZ) and W states under various decoherence channels with Kitagawa-Ueda (KU) criteria. In order to search spin squeezing sudden death (SSSD) and signatures of spin squeezing production we use bit flip, phase flip, bit-phase-flip, amplitude damping, phase damping and depolarization channels in the present study. In literature, the influence of decoherence has been studied as a destroying element. On the contrary here we investigate the positive aspect of decoherence, which produce spin squeezing in unsqueezed GHZ and W states under certain channels. Our meticulous study shows that GHZ state remain unsqueezed under aforementioned channels except bit-phase-flip and depolarization channels. While all the decoherence channels produce spin squeezing in W state. So we find, GHZ is more robust in comparison to W state in the sense of spin squeezing production under decoherence. Most importantly we find that none of the decoherence channel produce SSSD in any one of the state.
\end{abstract}
\pacs{1315, 9440T}
%\keywords{magnetic moment, solar neutrinos, astrophysics}
\maketitle
\section{\Large Introduction}
The phenomenon of squeezing\cite{sl1,sl2} exist in light from a long time in literature. Further the concept has been investigated in the context of spin systems by Kitagawa-Ueda (KU) in 1993\cite{mk,dw1,dw2}, so called spin squeezing. Spin squeezing attracted much attention of quantum information community as it has it's lucid connection with entanglement\cite{ent}. Many features of spin squeezing have been investigated to detect the multipartite and pairwise entanglement\cite{ment11}, even though negative pairwise entanglement\cite{entt} in quantum systems. A class of spin squeezing inequalities also established for the same purpose\cite{ment1,ment2,ment3,ment4,ment5}. Enhancement of sensitivity and precision in quantum metorology\cite{qm1,qm2} is an important aspect which has been achieved by using many techniques. Spin squeezing play an important role for quantum metrology and magnetometery to improve the sensitivity and precision because of it's ability to reduce the quantum noise\cite{qs1,qs2,qs3,qs4,qs5}. Early experimental manifestations of spin squeezing is demonstrated in varieties of
physical systems such as, in entangled ion trap systems\cite{ion}, in Bose Einstein condensate through repulsive interactions\cite{be}, measurements by partial projection\cite{pp}, squeezing of huge ensemble of atoms in cavities through light matter interaction to produce atomic clocks\cite{at1,at2,at3,at4,at5,at6,at7}, gravitational interferometers\cite{gw1,gw2,gw3} and in many qubit systems\cite{mq1,mq2,mq3}. There are many types of squeezing like dipolar, phase and planer squeezing including spin squeezing with various definitions\cite{mq5} and mathematical criteria. The definition depends on the type of the system and each has it's own significance. Here we deal with spin squeezing, which further demand any mathematical criteria to know it's degree. There are many criteria present for spin squeezing\cite{mq6}, KU is one of the highly studied criteria\cite{mk} which also have it's experimental manifestations. KU criteria has major drawback as it is mostly suitable for symmetric states and fail to check the spin squeezing under local unitary operations on the symmetric and non-symmetric states. However generalised version of KU criteria is developed by Usha Devi et al. for the states which are invariant under unitary transformations for non-symmetric as well as for symmetric states\cite{kc}. A well known example of symmetric states which are invariant under particle exchange is coherent spin states (CSS)\cite{css1,css2}. Primarily spin squeezing with KU criteria has been studied in CSS with one and two axis-twisting Hamiltonians, which are non-linear Hamiltonians and used to produce spin squeezing in separable CSS \cite{new1}. The best squeezing scales achieved with $N$ number of particles by both the Hamiltonians are $\epsilon_{OTH}^{2}\propto 1/N^{2/3}$ and $\epsilon_{TTH}^{2}\propto 1/N^{2}$ respectively without decoherence affects. The one axis twisting Hamiltonain under decoherence assumed in the form of particle loss, has been studied in two mode Bose-Einstein condensate and author have shown the spin squeezing production in the state\cite{ment11}. They have been observed, the scaling of spin squeezing with one particle loss is $\epsilon^{2}\propto N^{-4/15}$, for two particle loss the scaling is independent of $N$ and for three particle loss it is $\epsilon^{2}\propto N^{4/15}$. Further the spin squeezing production with two axis twisting model with decoherence in the form of particle loss has been studied by A. Andre et al. with Raman scattering based approach\cite{tth}. Early studies of decoherence in spin squeezing with CSS have been investigated under amplitude damping, pure dephasing and depolarizing channels by X. Wang et al\cite{sssd1}. This study have shown that CSS goes under spin squeezing sudden death (SSSD) under decoherence except amplitude damping channel. The phenomenon of SSSD is inspired by entanglement sudden death (ESD) which is widely studied in many physical settings\cite{ent1,ent2,ent3,ent4,ent5,ent6,ent7,ent8,ent9}. So investigations of spin squeezing behaviour under decoherence in CSS motivate us to carry out the study in another symmetric states. Here in the present study we consider tripartite maximally entangled GHZ and W states which are symmetric under particle exchange. The aim of the present study is two fold, we be diligent to search the phenomenon SSSD and some interesting results of spin squeezing production in these states under decoherence. 

The structure of the paper is as follows, in Sect.\ref{s2}, we give the brief formulation of spin squeezing, which is used to reckon the behaviour of spin squeezing through out the paper. In Sect.\ref{s3} and it's subsequent subsections from 3.1 to 3.6, we proceed the study on spin squeezing in tripartite GHZ and W states under decoherence channels like bit-flip, phase flip, bit-phase flip, amplitude damping, phase damping and depolarization. Lastly in Sect.\ref{s4} we provide the conclusion of the paper.
\section{\Large Spin Squeezing formulation}\label{s2}
In this section we cover basic definition of spin squeezing and it's formulation bases on KU criteria. In many body physics the collective behaviour of observables play an important role, because the individual particles can not be addressed like in Bose-Einstein-Condensates (BEC), in such systems spin squeezing play an important role to reduce the noise. Apart from it, spin squeezing also have strong connection with multipartite entanglement. To proceed the formulation of spin squeezing in N body spin-$\frac{1}{2}$ systems, we begin with the collective angular momentum $\vec{J}$. The  collective angular momentum $\vec{J}$ can be representing in three dimensional Cartesian coordinate system as below,
\begin{equation}
\vec{J}_{xyz}=(J_{x},J_{y},J_{z}).
\end{equation}
Where $J_{x}, J_{y}$ and $J_{z}$ are angular momentum components along x, y and z axis respectively. Further we transform the axis of euclidean space ie. $(x,y,z)$ to spherical coordinate system using the orthonormal vectors $(  
\vec{n}_{1},\vec{n}_{2},\vec{n}_{3})$. This transformation is nedded to remove the dipolar squeezing and to produce the spin squeezing in the system. The unit orthonormal vectors used for transformation are given below.
\begin{eqnarray}
\vec{n}_{1}=(-\sin \phi, \cos\phi, 0) \\
\vec{n}_{2}=(-\cos\theta \cos\phi,-\cos \theta \sin\phi,\sin \theta) \\
\vec{n}_{3}=(\sin\theta \cos\phi, \sin\theta \sin\phi,\cos\theta). \\
\end{eqnarray}
The collective angular momentum vector $\vec{J}_{xyz}$ in old coordinate system can be transformed in new coordinate system by using the rotation matrix as,
\begin{equation}
\left[
\begin{tabular}{c}
$J_{n_{1}}$ \\
$J_{n_{2}}$ \\
$J_{n_{3}}$
\end{tabular}
\right]=
\left[
\begin{tabular}{c c c c}
$-\sin\phi$ & $\cos\phi$ & $0$ \\
$-\cos\theta \cos\phi$ & $-\cos\theta \sin\phi$ & $\sin\theta$  \\
$\sin\theta \cos\phi$ & $\sin\theta \sin\phi$ & $\cos\theta$
\end{tabular}
\right].
\left[
\begin{tabular}{c}
$J_{x}$ \\
$J_{y}$\\
$J_{z}$
\end{tabular}
\right]. \label{trans}
\end{equation}
In L.H.S of the above equation, $(J_{n_{1}},J_{n_{2}},J_{n_{3}})$ are the components of the collective angular momentum $\vec{J_{n}}$ in new spherical coordinate system along the bases $( 
\vec{n}_{1},\vec{n}_{2},\vec{n}_{3})$.
Further simplification of equation (\ref{trans}) leads as,
\begin{eqnarray}
J_{n_{1}}=-J_{x}\sin \phi+J_{y}\cos \phi  \label{n1} \\
J_{n_{2}}=-J_{x}\cos \theta \cos \phi-J_{y}\cos \theta \sin \phi+J_{z}\sin \theta  \label{n2} \\
J_{n_{3}}=J_{x}\sin \theta \cos \phi+J_{y}\sin \theta \sin \phi+J_{z}\cos \theta. \label{n3}
\end{eqnarray}
The above equations \ref{n1}, \ref{n2} and \ref{n3} establish the connection of the components of vector $\vec{J}_{xyz}$ to the components of new vector $\vec{J}_{n}$ in spherical coordinate system. Knowing the components $(J_{x},J_{y},J_{z})$ and the set of angles $(\theta,\phi)$ gives the clue to obtain the components $(\vec{J}_{n_{1}},\vec{J}_{n_{2}},\vec{J}_{n_{3}})$ of the vector $\vec{J_{n}}$. In N spin-$\frac{1}{2}$ system, the angular momentum vectors of individual spins may be in different directions, so we need to calculate the mean values of the angular momentum components in x, y and z directions, these components form a mean spin vector $(\vec{J}_{mean})$ as obtained below,

\begin{equation}
\vec{J}_{mean}=\left(\langle J_{x} \rangle,\langle J_{y} \rangle, \langle J_{z} \rangle\right). \label{e1} \\
\end{equation}
The mean spin vector depends on the state of the system. Here it is important to mention that, we rotate the Cartesian coordinate system with the angles $(\theta, \phi)$ such that the mean spin vector $\vec{J}_{mean}$ align along the vector $\vec{n}_{3}$ in new spherical coordinate system, this makes calculations simple and assist the removal of dipolar spin squeezing in the system. Further, the mean spin vector in new spherical coordinate system can be expressed as
\begin{eqnarray}
(\vec{J}^{S}_{mean})=(\langle J_{n_{1}}\rangle,\langle J_{n_{2}}\rangle,\langle J_{n_{3}}\rangle).
\end{eqnarray}
with,
\begin{eqnarray}
\langle J_{n_{1}}\rangle= -\langle J_{x}\rangle \sin \phi+  
\langle J_{y}\rangle \cos \phi  \label{e2} \\
\langle J_{n_{2}}\rangle =
\langle -J_{x}\rangle \cos \theta \cos \phi-
\langle J_{y}\rangle \cos \theta \sin \phi+
\langle J_{z}\rangle \sin \theta  \label{e3} \\
\langle J_{n_{3}}\rangle =
\langle J_{x}\rangle \sin \theta \cos \phi+
\langle J_{y}\rangle \sin \theta \sin \phi+
\langle J_{z}\rangle \cos \theta. \label{e4}
\end{eqnarray}
Equations \ref{e2},\ref{e3} and \ref{e4} play an important role in calculating the spin squeezing, which basically revel the connections of mean components of the vector  $(\vec{J}^{S}_{mean})$ in spherical coordinate system to the mean components of the vector
\begin{figure*}[hbtp]
\centering
\includegraphics[scale=0.4]{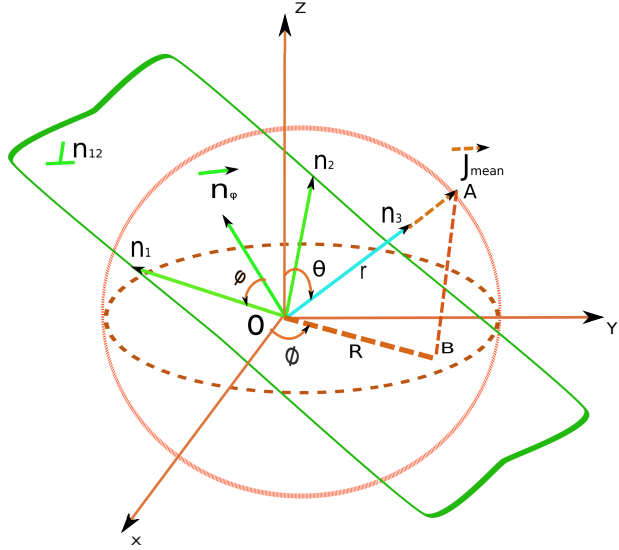} 
\caption{Arrangement of vectors $  
\vec{n}_{1},\vec{n}_{2},\vec{n}_{3}$,$\vec{J}_{mean}$, $\vec{J}_{\varphi}$ and plane $\vec{n}_{12}$}.\label{mf}
\end{figure*}
$\vec{J}_{mean}$ in Cartesian coordinate system. To go towards the mathematical definition of spin squeezing first we give the light on the arrangement of the vectors $\vec{n}_{1}$, $\vec{n}_{2}$ and $
\vec{n}_{3}$ which are orthonormal. Let assume the orthonormal vectors $\vec{n}_{1}$ and $\vec{n}_{2}$ are in the same plane denoted as $n_{12}$, obviously this plane is perpendicular to the vector $\vec{n}_{3}$ ie. $(n_{12}\perp \vec{n}_{3})$. We can choose any arbitrary vector lying in the plane $n_{12}$ and making an arbitrary angle $\varphi$ with the vector $\vec{n}_{1}$. This vector is called $\vec{n}_{\varphi}$, which is automatically perpendicular to the vector $\vec{n}_{3}$. The arrangement of the vectors $
\vec{n}_{1},\vec{n}_{2},\vec{n}_{3},\vec{n}_{\varphi}$, $\vec{J}_{mean}$ and $\vec{OB}$ is shown in the figure \ref{mf}. $\vec{OB}$ is the projection vector of $\vec{J}_{mean}$ in xy plane. As we know the vectors $\vec{n}_{1}$ and $\vec{n}_{2}$ are mutually perpendicular unit vectors which can be assumed as staying in two dimensional plane ie. $(\bot n_{12})$ and can be written as $\vec{n}_{1}=(1,0)$ and $\vec{n}_{2}=(0,1)$. Here we mention that rotation of the vector $\vec{n}_{1}$ with and angle $\varphi$ produce the vector $\vec{n}_{\perp}$. So $\vec{n}_{\perp}$ can be obtained as,
\begin{equation}
\vec{n}_{\perp}=S\vec{n}_{1} \label{nper}
\end{equation}
Where $S$ is the orthogonal rotation matrix expressed as,
\begin{equation}
S=\left[\begin{tabular}{c c}
$\cos \phi$ & $-\sin \phi$\\
$\sin \phi$ & $\cos \phi$
\end{tabular}\right]
\end{equation}
Simplification of equation \ref{nper} produce the vector $\vec{n}_{\perp}$, which is obtained below,
\begin{equation}
\vec{n}_{\varphi}=\vec{n}_{1}\cos \varphi+\vec{n}_{2}\sin \varphi.
\end{equation}

The vector $\vec{n}_{\varphi}$ lies in the plane $n_{12}$ in spherical coordinate system and play an important role for the definition of spin squeezing. The definition of spin squeezing is given as \textit{``The minimum value of the variance of the total angular momentum $J_{n}$ along the vector $\vec{n}_{\varphi}$ is less than or equal to the stranded quantum limit"}. ie. 
\begin{equation}
[(\vartriangle J_{\varphi})^{2}]_{min}\leq\frac{J}{2}. \label{e7} 
\end{equation} 
Where $J$ is the spin quantum number and given as $J=\frac{N}{2}$. Rearranging the equation (\ref{e7}) we get
\begin{equation}
\epsilon=\frac{[4(\vartriangle J_{\varphi})^{2}]_{min}}{N}\leq 1. \label{e8}
\end{equation}
Where $N$ is the number of spins in the system and $\epsilon$ is the spin squeezing parameter. If $(\epsilon=1)$, the state is unsqueezed and there is no quantum correlation present in the state. For pure CSS uncorrelated states $(\epsilon=1)$. If there are certain type of quantum correlations present in the state, than $(\epsilon\leq 1)$.
We can easily obtain the projection of the vector $\vec{J}_{n}$ along the vector $\vec{n}_{3}$, which produce the vector $\vec{ J}_{\varphi}$. This vector is obtained as,
\begin{equation}
\vec{J}_{\varphi}=\vec{J}_{n}.\vec{n}_{\varphi}=J_{n_{1}}\cos\varphi+J_{n_{2}}\sin\varphi. \label{e5}
\end{equation} 
The variance of the vector $\vec{J}_{\varphi}$ and it's optimization is obtained in appendix \ref{ap1}, which is expressed as,
\begin{equation}
(\vartriangle J_{\varphi})^{2}_{min}=\frac{1}{2}[O- \sqrt{M^{2}+N^{2}}].\label{tj}
\end{equation}
Where the coefficients $M$, $N$ and $O$ are given below,
\begin{eqnarray}
M=\langle J_{n_{1}}^{2}-J_{n_{2}}^{2}\rangle \\
N=\langle J_{n_{1}}J_{n_{2}}+J_{n_{2}}J_{n_{1}}\rangle \\
O=\langle J_{n1}^{2}+J_{n_{2}}^{2}\rangle. \\
\end{eqnarray}

Putting the value from the equation (\ref{tj}) in equation (\ref{e8}) we get the spin squeezing as,
\begin{equation}
\epsilon=\frac{2}{N}(O-\sqrt{M^{2}+N^{2}})\leq 1. \label{ssp}
\end{equation}
This is called KU criteria for spin squeezing. The factor $(\epsilon < 1)$ represents the state is spin squeezed and with $(\epsilon =1)$ the state is unsqueezed. However the criteria also gives the indications about the separability and entanglement of a state. The parameter $(\epsilon \leq 1)$ exhibits that the state is entangled and  with $(\epsilon= 1)$, the state is separable. But this is true only for symmetric states, the criteria fails for non symmetric states and under local operations. 
\section{\Large Spin Squeezing under decoherence in GHZ and W states}\label{s3}
In this section we study the spin squeezing behaviour in tripartite maximally entangled GHZ and W states under decoherence. We here consider that individual qubits face the decoherence, this situation is modelled by Kraus operators formalism \cite{kr}. 
To start with, the tripartite maximally entangled GHZ and W states can be written as,
\begin{eqnarray}
|\psi\rangle_{GHZ}=\frac{1}{\sqrt{2}}(|000\rangle+|111\rangle). \label{ghz} \\
|\psi\rangle_{W}=\frac{1}{\sqrt{3}}(|100\rangle+|010\rangle+|001\rangle). \label{ghz}
\end{eqnarray}
Both the states are symmetric under particle exchange. The corresponding initially density matrices of these states can be obtained as
$\rho_{GHZ}(0)$ and $\rho_{W}(0)$. In subsequent subsections we proceed the study under bit flip, phase flip, bit-phase flip, amplitude damping, phase damping and depolarization channels. Let initial density matrix passes through any one of the quantum channel, than by using Kraus operators the decoherence prone density matrix can be written as,
\begin{equation}
\rho^{dp}=\sum_{i=1}^{n}\sum_{j=1}^{n}\sum_{k=1}^{n}E_{i}\otimes E_{j}\otimes E_{k}[\rho(0)]E_{i}^{\dagger}\otimes E_{j}^{\dagger}\otimes E_{k}^{\dagger}. \label{dec}
\end{equation}
Where $n$ is the number of kraus operators for a particular channel and $E_{i,j,k}$ are the Kraus operators for three qubits $i,j$ and $k$, which are operating independently.  The state $\rho(0)$ represents the initial state either $\rho_{GHZ}(0)$ or $\rho_{W}(0)$. The equation (\ref{dec}) will be used for the decoherence analysis in spin squeezing throughout the paper for various quantum channels.
\subsection{Bit flip channel:}
In this subsection we study the spin squeezing behaviour of GHZ and W states under bit flip channel. Bit flip error is a common error produced in many physical systems, a common example is the stray magnetic field which may flip the bit and produces this error. To begin with we write the Kraus operators of bit flip channel, which are given as,
\begin{equation}
E_{1}=
\left[\begin{tabular}{c c}
$\sqrt{p}$ & $0$\\
$0$ & $\sqrt{p}$
\end{tabular}\right]\\
E_{2}=\left[\begin{tabular}{c c}
$0$ & $\sqrt{(1-p)}$\\
$\sqrt{(1-p)}$ & $0$
\end{tabular}\right].
\end{equation}
The bit flip channel has two Kraus operators, we plug-in these operators with $n=2$, in equation (\ref{dec}). Putting   $\rho(0)=\rho_{GHZ}(0)$ for GHZ state and $\rho(0)=\rho_{W}(0)$ for W state in equation (\ref{dec}), we get the decoherence prone density matrix $\rho^{dp}_{bf}$ for bit flip channel, which is used to reckon the spin squeezing in GHZ and W states by using equation (\ref{ssp}). The expressions of degree of spin squeezing in GHZ and W states are obtained as,
\begin{dmath}
\epsilon_{GHZ}=\frac{1}{2} \left[-2(2 p-1)^2 \sin ^2(\theta )-(1-2 p)^2\cos (2 \theta)+4 (p-1) p+3\right].\label{bfg}
\end{dmath}
\begin{dmath}
\epsilon_{W}=
\frac{2}{3}\left [-\sin(\theta)\cos(\theta)\cos(\phi)+\frac{3\cos^2(\theta)}{8}+\left(2(p-1)p+\frac{7}{8}\right)\sin ^2(\theta)+p\sin(2\theta)\cos(\phi)-(2p-1)\sin(\theta)\sqrt{\sin^2(\phi)+\frac{1}{4}
[2\cos (\theta)\cos(\phi)+(2p-1)\sin(\theta)]^2}+\frac{9}{8}\right ].\label{wbf}
\end{dmath}
First, we reckon the behaviour of spin squeezing in GHZ state with the equation (\ref{bfg}). In the absence of bit flip error ie. $(p=0)$, the  equation (\ref{bfg}) become free from the angle $\theta$ and we get $(\epsilon_{GHZ}=1)$, which revels the state is initially unsqueezed. Taking the other side of the discussion we can also look at the length of mean spin vector in GHZ state, which is zero
in the absence of decoherence. We calculate the length of the mean spin vector $(r)$ for GHZ state in the decoherence prone matrix $\rho^{dp}_{bf}$ by using the equation (\ref{rr}), which  is still found as zero and independent from the parameter $p$. So, the result $(r=0)$ in the state $\rho^{dp}_{bf}$ represents the origin $``O"$ in the figure \ref{mf}, which revels that $(\theta,\phi)=(0,0)$ in GHZ state. So by putting $(\theta=0)$ in equation (\ref{bfg}) we get,
\begin{eqnarray}
\epsilon_{GHZ}=\frac{1}{2}\left[-(1-2p)^{2} +4(p-1)p+3\right]
=1.
\end{eqnarray}
The parameter $(p)$ vanish from the equation and we get $(\epsilon_{GHZ}=1)$. This concludes that the GHZ state remains unsqueezed and do not feel the influence of bit flip channel and of course do not exhibit the signatures of spin squeezing production.
\begin{figure*}[hbtp]
\centering
\includegraphics[scale=1]{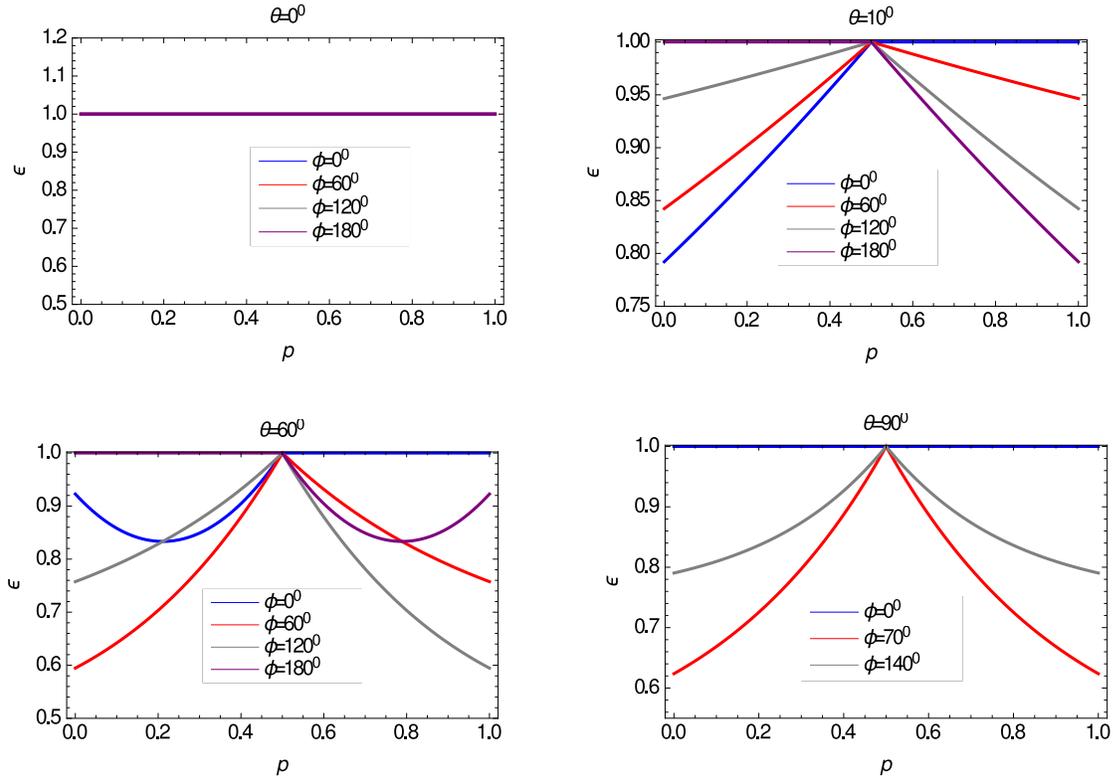}
\caption{Plot of $\epsilon_{W}$ vs. parameter $p$ with $\phi\in [0^{0},180^{0}]$ for bit flip channel} \label{bf}
\end{figure*}
Next we give the look at equation (\ref{wbf}), in which the spin squeezing parameter $\epsilon_{W}$ is the function of three parameters $(\theta,\phi,p)$. We plot $\epsilon_{W}$ vs. parameter $p$ with different values of $(\theta,\phi)$ in figure \ref{bf}. It is found that in the absence of bit flip error ie. $(p=0)$, the length of the mean spin vector in W state is $(r=0.372678\neq 0)$, which is reverse case to the GHZ sate. So there is a possibility for spin squeezing and it's production in W state. Giving the look at the subfigure of figure \ref{bf} with $(\theta=0^{0})$, we find with $p\in [0,1]$ and $\phi \in [0^{0},180^{0}]$, the parameter $\epsilon_{W}$ is always equal to 1. Which concludes that as long as the mean spin vector is along the z-axis, the state remain unsqueezed and the state remain unaffected by bit flip channel. But as the mean spin vector rotates with $\theta \in (0^{0}, 90^{0}]$ and $\phi\in [0^{0}, 180^{0}]$ the spin squeezing produce in the state. Direction of the mean spin vector has important significance which play the role to determine the plane of reduced variances, which lie normal to the direction of mean spin vector. By giving the look at the figure \ref{bf} with $(\theta=10^{0},60^{0},90^{0})$, we find the behaviour of spin squeezing parameter $\epsilon_{W}$,
such that the bit flip parameter $p$ and rotation angle $\phi$ produce the spin squeezing in the state. But it is important to note that with $(p=0.5)$, the parameter $(\epsilon_{W}=1)$, and there is no spin squeezing produced. Here we mention that the movement of mean spin direction represents the movement of projection vector $\vec{OB}$ in xy plane  with an angle $\phi$ and vice versa. With $p\in [0,0.5],\theta \in (0,90^{0}]$ as the projection vector $\vec{OB}$ rotates in the xy plane with an angle $\phi$ the spin squeezing parameter achieves the value as $(\epsilon_{W}<1)$, so spin squeezing signatures are produced. The condition $[\phi=0^{0},\theta \in (0^{0},90^{0})]$ represents that the mean spin lie in xz plane and $[\phi=180^{0},\theta \in (0^{0},90^{0})]$ represents that the mean spin lie in the plane yz. As mean spin or projection vector $\vec{OB}$ rotates with $\phi\in (0^{0},180^{0})$ in xy plane, the degree of spin squeezing rises with $(p<0.5)$ and decreases with $(p>0.5)$. 
Bases on discussion for the results obtained in figure \ref{bf}, we conclude that the W state is fragile under bit flip channel and shows the signatures of spin squeezing production.
\subsection{Phase flip channel}
In this section we study the spin squeezing behaviour of GHZ and W states under phase flip channel. The Kraus operators of phase flip channel are given below.
\begin{equation}
E_{1}=
\left[\begin{tabular}{c c}
$\sqrt{p}$ & $0$\\
$0$ & $\sqrt{p}$
\end{tabular}\right]\\
E_{2}=\left[\begin{tabular}{c c}
$\sqrt{(1-p)}$ & $0$ \\
$0$ & $\sqrt{(1-p)}$ 
\end{tabular}\right].
\end{equation}
By using the equation (\ref{dec}) with $n=2$, we can obtain the decoherence prone matrix corresponding to the phase-flip channel as $\rho^{dp}_{fp}$. Further calculating the spin squeezing parameter in GHZ and W states by using the equation (\ref{ssp}) we get.
\begin{dmath}
\epsilon_{GHZ}=
\frac{1}{2}\left(-2\sin^2(\theta)-\cos(2\theta)+3\right)=1.\label{pfghz}
\end{dmath}
\begin{dmath}
\epsilon_{W}=\frac{2}{3}\left[
\frac{1}{4}
\{-\cos(2\theta)+2(2p-1)\sin(2\theta)\cos(\phi)+7\}
-\sqrt{(2p-1)^2\sin^2(\theta)\sin^2(\phi)+\frac{1}{4}\{(1-2p)\sin(2\theta)\cos(\phi)-\sin^2(\theta)\}^2}\right].\label{pfwe}
\end{dmath}
Equation \ref{pfghz} reveals, the equation is free from the decoherence parameter p and simplification of the equation (\ref{pfghz}) leads as $(\epsilon_{GHZ}=1)$. In fact this equation is similar to equation (\ref{bfg}) with $(p=0)$. With $(\epsilon_{GHZ}=1)$, we conclude that GHZ state do not feel decoherence by phase flip channel and avoid the signatures of spin squeezing production.

Now we concentrate at the squeezing parameter obtained for W state in equation (\ref{pfwe}). The squeezing parameter $\epsilon_{W}$ is the function of three parameters $(\theta,\phi,p)$. 
\begin{figure*}[hbtp]
\centering
\includegraphics[scale=0.7]{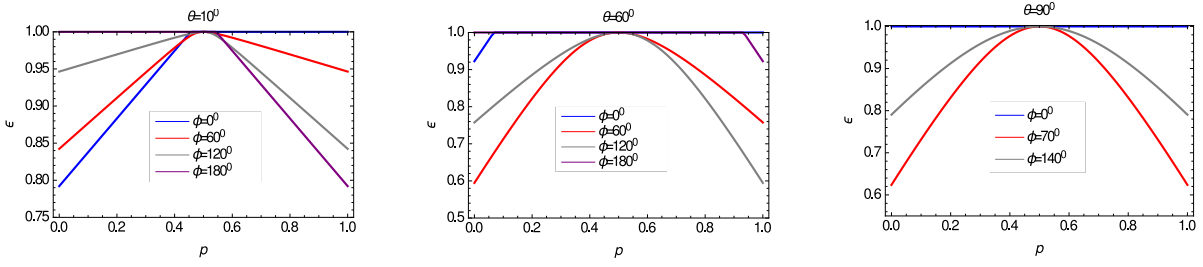}
\caption{Plot of parameter $\epsilon_{W}$ vs. parameter $p$ with $\phi\in [0^{0},180^{0}]$ for phase flip channel} \label{fp}
\end{figure*}
With varying values of these parameters the results are plotted in figure \ref{fp}. At $(\theta=0^{0})$ under phase flip channel, we obtain the similar result as obtained with $(\theta=0^{0})$ for bit flip channel as shown in figure \ref{bf}. This result shows that as long as the mean spin vector is along the z axis, the W state remains initially unsqueezed and unaffected by phase flip channel. Here in figure \ref{fp} with $\theta\in (0^{0},90^{0}]$, spin squeezing takes round in the vicinity of $(p=0.5)$. However for bit flip channel the spin squeezing was sharply meeting at $(p=0.5)$. At $(p=0)$ and $(p=1)$, the degree of spin squeezing obtained under this channel is similar to bit flip channel with varying values of $(\theta,\phi)$. When mean spin vector is in xz plane with $(\theta=60^{0},\phi=0^{0})$, the spin squeezing is produced only with $p\in [0.05,0.1]$. After $(p>0.1)$, the state is unsqueezed. Once the mean spin vector switch to yz plane with $(\theta=60^{0},\phi=180^{0})$, the state is squeezed only with $p\in[0.95,1.0]$. As the projection vector $\vec{OB}$ rotates in xy plane, except $(\phi=0^{0},\phi=90^{0})$, there are good features of spin squeezing production. Here we mention that phase flip channel has the capability to produce spin squeezing in W sate under decoherence.

\subsection{Bit-Phase-Flip channel}
\begin{figure*}[hbtp]
\centering
\includegraphics[scale=1.5]{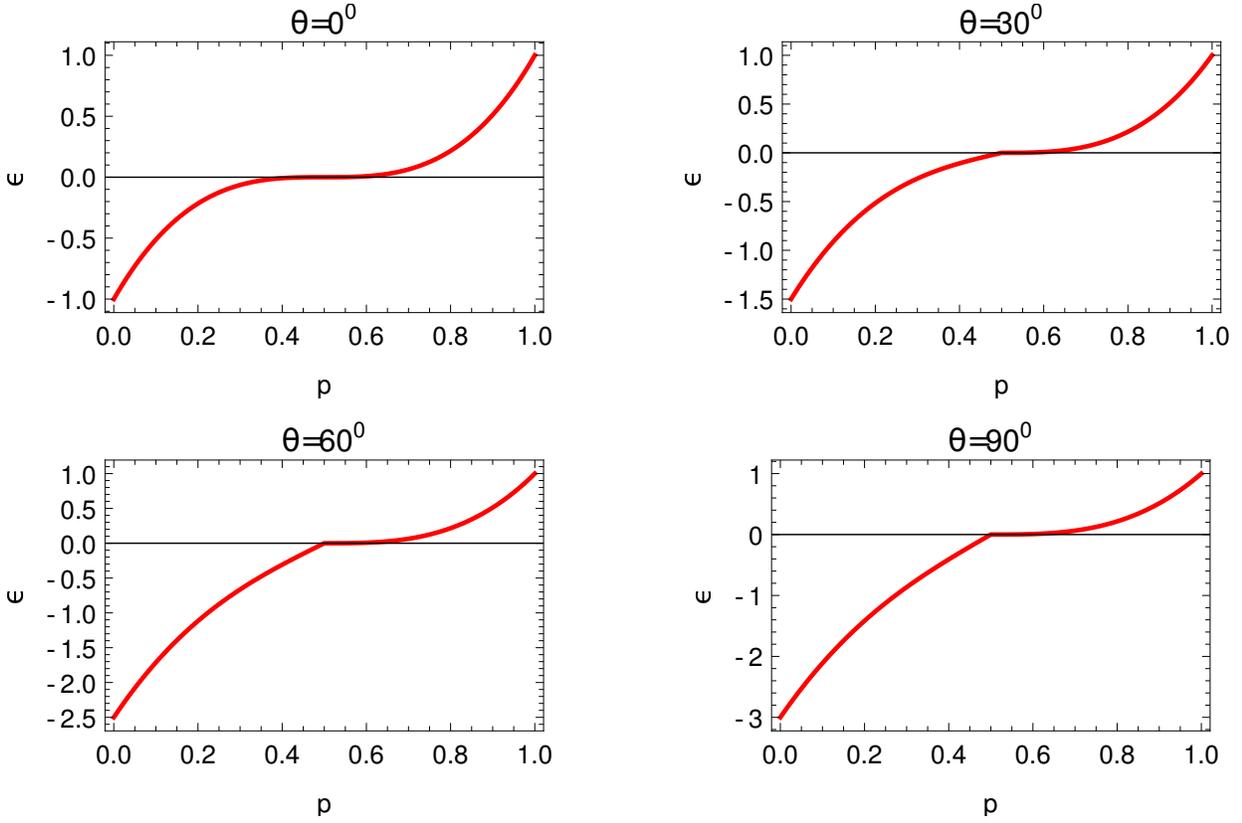}
\caption{Spin squeezing Plot for bit-phase-flip in GHZ state} \label{bfpf1}
\end{figure*}

In this section we study the behaviour of spin squeezing under bit-phase-flip channel. This channel flip the bit along with the emergence of relative phase factor in the state. The kraus operators of bit-phase-flip channel are given below.
\begin{equation}
E_{1}=
\left[\begin{tabular}{c c}
$\sqrt{p}$ & $0$\\
$0$ & $\sqrt{p}$
\end{tabular}\right]\\
E_{2}=\left[\begin{tabular}{c c}
$0$ & $-i\sqrt{(1-p)}$\\
$i\sqrt{(1-p)}$  & $0$ 
\end{tabular}\right].
\end{equation}
We use the equation (\ref{dec}) with $n=2$, and obtain the decoherence prone density matrix after passing through the channel as $\rho^{dp}_{bfp}$. By using $\rho^{dp}_{bfp}$ The spin squeezing parameter for GHZ and W states are obtained as,
\begin{dmath}
\epsilon_{GHZ}=
\frac{1}{2}\left[(2 p-1)(-\cos(2\theta)+8(p-1)p+3)-2(1-2p)\sin^2(\theta)\right]. \label{gbfp}
\end{dmath}
\begin{dmath}
\epsilon_{W}=
\frac{2}{3}\left[\frac{1}{4}(2 p-1)\{2\sin(2\theta)\cos(\phi)-\cos(2\theta)+24(p-1)p+7\}-(2 p-1)\sqrt{\sin^2(\theta)\sin^2(\phi )+\frac{1}{4}\{\sin(2\theta)\cos(\phi)
+\sin^2(\theta)\}^2}\right].\label{wbfp}
\end{dmath}
Looking at the equation \ref{gbfp}, in the absence of decoherence i.e. $(p=0)$, we get $(\epsilon_{GHZ}=1)$, hence the state  is initially  unsqueezed. As the direction of mean spin vector ie. $(\theta)$ and the value of parameter $p$ increases, the state become squeezed. The result is shown in figure \ref{bfpf1}. 
We also observe in the figure \ref{bfpf1} with $(p\leq 0.5)$, the squeezing parameter achieve negative values, which is an indicator that the channel induces negative correlations in the state\cite{ment2}. With $(p>0.5)$, as the value of parameter $p$ increases the degree of spin squeezing exponentially increases. Here we conclude, the bit-phase-flip channel has the capability to produce the signatures of spin squeezing in GHZ state.

\begin{figure*}[hbtp]
\centering
\includegraphics[scale=1]{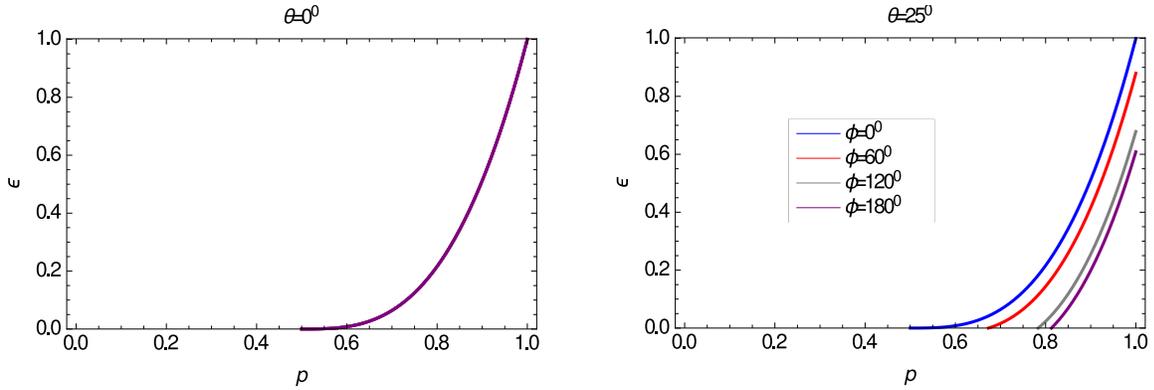}
\caption{Plot of squeezing paraneter $\epsilon_{W}$ vs. parameter $p$ with $\phi\in[0^{0},180^{0}]$}. \label{bpf}
\end{figure*}

Further for W state, giving the glance to equation (\ref{wbfp}), we find the squeezing parameter is the function of three parameters $(\theta,\phi,p)$. The results based on this equation are plotted in figure \ref{bpf}. With $(\theta=0^{0})$, the equation becomes free from the angle $(\phi)$ and remains the function of parameter $(p)$. With $(\theta=0^{0})$, the mean spin vector in this case is along the z axis and result is plotted in figure \ref{bpf}. Which reveals that as the value of probability increases with $(p>0.5)$, the degree of spin squeezing exponentially increases. But there is no spin squeezing produced in the state with $(p<0.5)$ . Further observing the figure \ref{bpf} with $(\theta=25^{0})$, reveals that as the direction of mean spin vector changes, the degree of spin squeezing exponentially grows with varying values of $(\phi)$. As the value of the angle $(\phi)$ changes, the squeezing is produced with higher values of $(p)$ beyond the range $(p>0.5)$.
Here we find, the bit-phase-flip channel produces  spin squeezing  signatures with $(p>0.5)$ in both the GHZ and W states. So both the GHZ and W sates exhibit fragile behaviour 
with bit-phase-flip channel in the sense of spin squeezing production.
\subsection{Amplitude damping channel}
In this section, we study the spin squeezing under amplitude damping channel. This channel is used to  describe the dissipation of the interaction between quantum systems and it's environment, for example the spontaneous emission of the photon from a quantum system. In actual the amplitude damping channel describe the energy loss in the system. The kraus operators of amplitude damping channel is given below,
\begin{equation}
E_{1}=
\left[\begin{tabular}{c c}
$1$ & $0$\\
$0$ & $\sqrt{e^{-\gamma t}}$
\end{tabular}\right]\\
E_{2}=\left[\begin{tabular}{c c}
$0$ & $\sqrt{1-e^{-\gamma t}}$\\
$0$ & $0$ 
\end{tabular}\right].
\end{equation}
Where $\gamma$ is the damping rate for the channel. First, we obtain the density matrix corresponding to the amplitude damping channel by using the equation (\ref{dec}) with $n=2$. After that spin squeezing parameters for GHZ and W states are obtained as below,
\begin{dmath}
\epsilon_{GHZ}=
\frac{1}{2}\left[-2e^{-2\gamma t}\left(e^{\gamma t}\left(e^{\gamma t}-2\right)+2\right)\sin^2(\theta )-4e^{-2\gamma t}\left(e^{\gamma t}-1\right)\sin^2(\theta)-\cos(2\theta)+3\right]
=1. \label{adghz}
\end{dmath}
\begin{dmath}
\epsilon_{W}=
\frac{2}{3}\left[\frac{1}{4}\left(8\left(e^{-\gamma t}\right)^{3/2}\sin(\theta)\cos(\theta)\cos(\phi)-2
\sqrt{e^{-\gamma t}}\sin(2\theta)\cos(\phi)- \\
4e^{-2\gamma t}\left(3 e^{\gamma t}-2\right)\sin^2(\theta)-3 \cos (2 \theta )+9\right)-\sqrt{A_{1}}\right].\label{wad}
\end{dmath}
with
\begin{dmath}
A_{1}=e^{-3\gamma t} \left(e^{\gamma t}-2\right)^2\sin^2(\theta)\sin^2(\phi)+\frac{1}{4}e^{-3\gamma t} \sin^2(\theta ) \left(\sqrt{e^{-\gamma t}}\left(3e^{\gamma t}\left(e^{\gamma t}-2\right)+4\right) \sin (\theta)\\
-2 \left(e^{\gamma t}-2\right) \cos(\theta)\cos(\phi)\right)^2.
\end{dmath}
Simplification of the equation (\ref{adghz}) leads as $(\epsilon_{GHZ}=1)$. We find, the squeezing parameter $\epsilon_{GHZ}$ is independent from decoherence parameter $(\gamma t)$ and angle $(\theta)$. So it implies that GHZ state do not feel the decoherence  from amplitude damping channel and exhibit the robust character against the spin squeezing production.

Equation (\ref{wad}) is used to explore the results for W state. 
\begin{figure}[hbtp]
\centering
\includegraphics[scale=0.7]{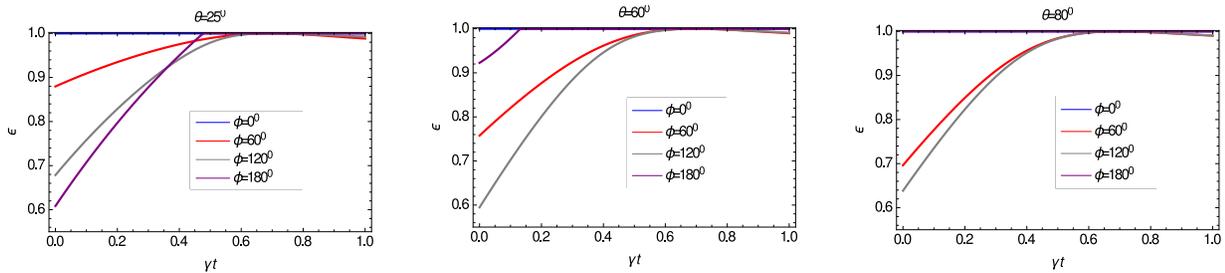}
\caption{Spin squeezing Plot amplitude damping channel.} \label{adc}
\end{figure}
The results are shown in figure \ref{adc}. When mean spin vector is along the z axis with  $(\theta=0^{0})$, the equation (\ref{wad}) leads the value $(\epsilon_{W}=1)$, which represents the state is unsqueezed.  As the value of the parameter $(\gamma t,\theta)$ increases with variations in the angle $(\phi)$, the spin squeezing is produced in the state. Further we find there are good features of spin squeezing production with $(\gamma t<0.6)$. It is found, with $(\phi=0^{0},\forall \theta)$, the state is always unsqueezed. We notice, as the mean spin vector rotates over the plane xy with $\phi\in (0^{0},180^{0})$, there are nice signatures of spin squeezing production. These are clearly shown in figure \ref{adc} with increasing values of parameter $(\theta)$. Here we find, W state shows fragile character under the specified channel.
\subsection{Phase damping channel}
In this section we study the spin squeezing behaviour under phase damping channel. Phase damping channel is the model to represent the information loss in quantum system because of the relative phase produced in the system with system enviornment interaction. This channel do not involve the energy loss in the system as it is done in the case of amplitude damping channel. Recently it is observed that spin squeezing can be produced  with phase damping channel in the system by using quantum non demolition interaction  (QND)\cite{qnd1,qnd2,qnd3,qnd4,qnd5,qnd6,qnd7,qnd8}, so it is important to study the affect of this channel on spin squeezing. The kraus operators are given for this channels as below.

\begin{equation}
E_{1}=
\left[\begin{tabular}{c c}
$\sqrt{e^{-\gamma t}}$ & $0$\\
$0$ & $\sqrt{e^{-\gamma t}}$
\end{tabular}\right]\\
E_{2}=\left[\begin{tabular}{c c}
$\sqrt{1-e^{-\gamma t}}$ & $0$\\
$0$ & $0$ 
\end{tabular}\right]
E_{3}=\left[\begin{tabular}{c c}
$0$ & $0$ \\
$0$ & $\sqrt{1-e^{-\gamma t}}$ 
\end{tabular}\right].
\end{equation}

We obtained the density matrix $\rho^{dp}_{pdc}$ by putting $n=3$ in equation (\ref{dec}). Further we have calculated the spin squeezing parameters for phase damping channel for both the GHZ and W states, these are obtained below,
\begin{figure}[hbtp]
\centering
\includegraphics[scale=0.7]{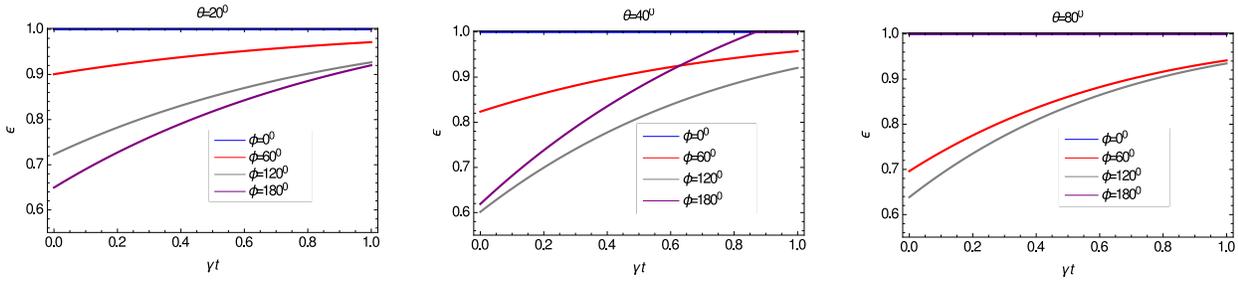}
\caption{Spin squeezing Plot phase damping channel.} \label{phdam}
\end{figure}
\begin{dmath}
\epsilon_{GHZ}=\frac{1}{2}[-2\sin^2(\theta)-\cos(2\theta )+3]=1\label{pdghz}
\end{dmath}
\begin{dmath}
\epsilon_{W}=
\frac{2}{3}\left[e^{-\gamma t}\sin(\theta)\cos(\theta)\cos(\phi)-\sqrt{e^{-2\gamma t}\sin^2(\theta)
\sin^2(\phi)-\left(A_{2}\right)^2}-\frac{1}{4}\cos(2\theta)+\frac{7}{4}\right].\label{pdw}
\end{dmath}
with 
\begin{dmath}
A_{2}=\sin(\theta)\cos(\theta)\cos(\phi)(e^{-\gamma t})+\frac{\sin^2(\theta)}{2}.
\end{dmath}
Simplification of equation (\ref{pdghz}) shows that for GHZ state the spin squeezing parameter is obtained as $(\epsilon_{GHZ}=1)$, it represents the state is unsqueezed and channel do not produce spin squeezing in the state. 

Further for W state, with the equation (\ref{pdw}) we find at $(\theta=0^{0})$, the squeezing parameter is $(\epsilon_{W}=1)$. So  it implies as long as the mean spin vector is along the z axis, the W state is unsqueezed. For higher values of the parameter $(\theta)$, the results are plotted in figure \ref{phdam}. We found, as the mean spin vector is in xz plane with $(\phi=0^{0},\forall \theta)$, the spin squeezing  
has not been produced in the state. While the rotation of mean spin vector with $\phi\in (0^{0},180^{0})$ produce spin squeezing signatures in the state with the increasing values of decoherence parameter $(\gamma t)$.
\subsection{Depolarization channel}
Under this section we study the spin squeezing  behaviour under depolarization channel. This channel is widely studied in polarization encoding in quantum information, the map of depolarization is described as it lives the system in fully mixed state with the probability $(p)$ and the systems is unchanged with the probability $(1-p)$. The kraus operators for depolarization channels are given below. 
\begin{equation}
E_{1}=
\left[\begin{tabular}{c c}
$\sqrt{e^{-\gamma t}}$ & $0$\\
$0$ & $\sqrt{e^{-\gamma t}}$
\end{tabular}\right]\\
E_{2}=\left[\begin{tabular}{c c}
$0$ & $\sqrt{\frac{1}{3}(1-e^{-\gamma t})}$ \\
$\sqrt{\frac{1}{3}(1-e^{-\gamma t})}$ & $0$ 
\end{tabular}\right].
\end{equation} 

\begin{equation}
E_{3}=
\left[\begin{tabular}{c c}
$0$ & $-i\sqrt{\frac{1}{3}(1-e^{-\gamma t})}$ \\
$i\sqrt{\frac{1}{3}(1-e^{-\gamma t})}$ & $0$
\end{tabular}\right]\\
E_{4}=\left[\begin{tabular}{c c}
$\sqrt{\frac{1}{3}(1-e^{-\gamma t})}$  & $0$ \\
$0$ & $-\sqrt{\frac{1}{3}(1-e^{-\gamma t})}$
\end{tabular}\right]. 
\end{equation}
\begin{figure*}[hbtp]
\centering
\includegraphics[scale=1.0]{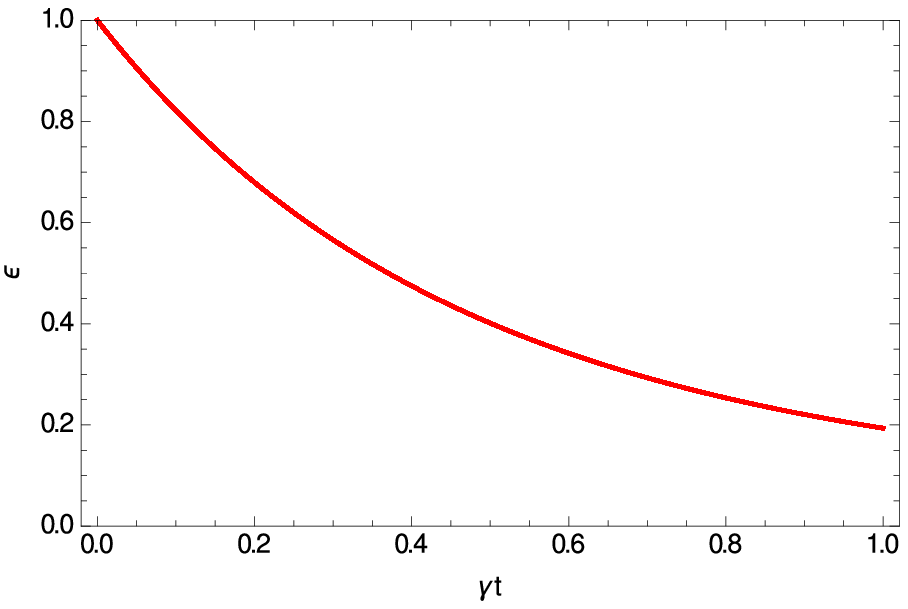}
\caption{Spin squeezing plot for GHZ state under Depolarizing noise.} \label{dpn}
\end{figure*}
We obtained the density matrix $\rho^{dp}_{pdc}$ by putting $n=4$ in equation (\ref{dec}). Further, we obtained the spin squeezing parameters for GHZ and W states with $\rho^{dp}_{pdc}$. These are obtained below,
\begin{dmath}
\epsilon_{GHZ}=
\frac{1}{54}\left(-2e^{-3\gamma t} \left(e^{\gamma t}+2\right)^3\sin^2(\theta)-e^{-3\gamma t}\left(e^{\gamma t}+2\right)^3(\cos(2\theta)-3)\right)\\
=\frac{1}{27}e^{-3\gamma t}(e^{\gamma t}+2)^{3}.\label{ddghz}
\end{dmath}
\begin{dmath}
\epsilon_{W}=
\frac{2}{3}\left[-\sqrt{\frac{1}{729}e^{-6\gamma t}\left(e^{\gamma t}+2\right)^6 \sin^2(\theta ) \sin ^2(\phi)+\frac{A_{3}}{2916}}-\frac{1}{108} e^{-3 \gamma t}\left(e^{\gamma t}+2\right)^3(-2\sin(2\theta)\cos(\phi)+\cos(2\theta)-7)\right].\label{ddw}
\end{dmath}
with
\begin{dmath}
A_{3}=e^{-6 \gamma t}\left(e^{\gamma t}+2\right)^6 \sin^2(\theta)(2\cos(\theta)\cos(\phi )+\sin (\theta))^2.
\end{dmath}

Looking at the equation \ref{ddghz}, we find the squeezing parameter for GHZ state is the function of the damping rate $(\gamma t)$ and independent from the parameters $(\theta,\phi)$. It implies that, the mean spin can be in any direction in the space.
We have plotted this function in figure \ref{dpn}.
At $(\gamma t=0)$ we have $(\epsilon_{GHZ}=1)$, the state is initially unsqueezed, but as the depolarization rate increases the spin squeezing is produced in the state, which decay exponentially and stay at $\epsilon_{GHZ}=0.2$. So this channel shows lucid signatures for spin squeezing production in GHZ state and the state is very much fragile under this channel.
\begin{figure*}[hbtp]
\centering
\includegraphics[scale=0.7]{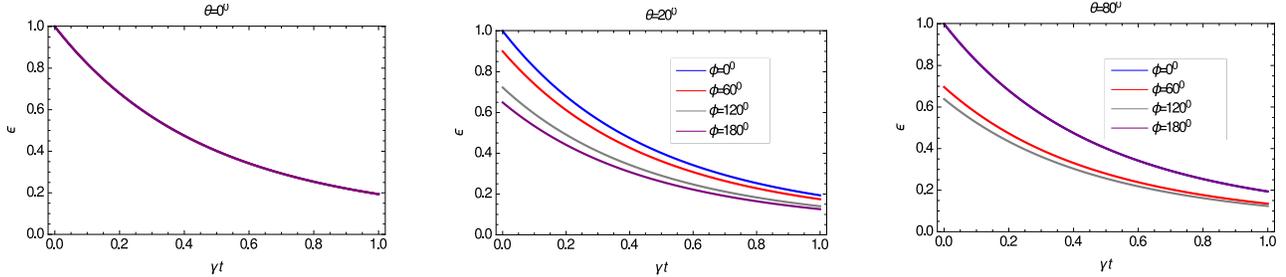} 
\caption{Spin squeezing Plot for GHZ state under Depolarizing noise.} \label{dp2n}
\end{figure*}

For W state, we study the equation (\ref{ddw}), we find at $(\theta=0^{0})$, the equation convert into the equation (\ref{ddghz}) and spin squeezing in W state exhibit the same  behaviour as found in GHZ state. This result is shown in figure \ref{dp2n} with $(\theta=0^{0})$. Looking at subfigures of figure \ref{dp2n}, we find as the values of $(\theta)$ increases the spin squeezing produces in W state and decreases exponentially as the depolarization rate $(\gamma t)$ increases. Most importantly we have found, as the mean spin vector lies in xz or in yz plane with $(\phi=0^{0},\phi=180^{0},\forall \theta)$, still the spin squeezing is produced in W state under depolarization channel. We have found the depolarization channel has great capability to produce spin squeezing in both the GHZ and W states. \section{\Large Conclusion}\label{s4}
In this article we investigate the behaviour of spin squeezing in tripartite maximally entangled GHZ and W states under bit flip, phase flip, bit-phase flip, amplitude damping, phase damping and depolarization channels. Initially 
GHZ state is unsqueezed and W state is also unsqueezed as long as it's mean spin vector is along the z axis. When 
decoherence is applied, we have found the lucid signatures of spin squeezing production in these states. However we have found that GHZ state remain unsqueezed under all the decoherence channels  except bit-phase-flip and depolarization channels. The W state shows fragile behaviour under decoherence and it permits all the channels to produce spin squeezing in it and exhibit less robust character than GHZ state. More specifically we have found depolarization channel has lucid characteristic to produce good degree of spin squeezing in both the GHZ and W states. We also have investigated that none of the state exhibit spin squeezing sudden death under any one of the decoherence channel. Investigating the positive aspect of decoherence on spin squeezing in tripartite GHZ and W states can be a useful study in quantum information processing.

\appendix
\section{\Large Calculations of variance $(\vartriangle J_{\varphi})^{2}$}\label{ap1}
\begin{figure*}[hbtp]
  \centering
  \begin{minipage}[b]{0.35\textwidth}
    \includegraphics[width=\textwidth]{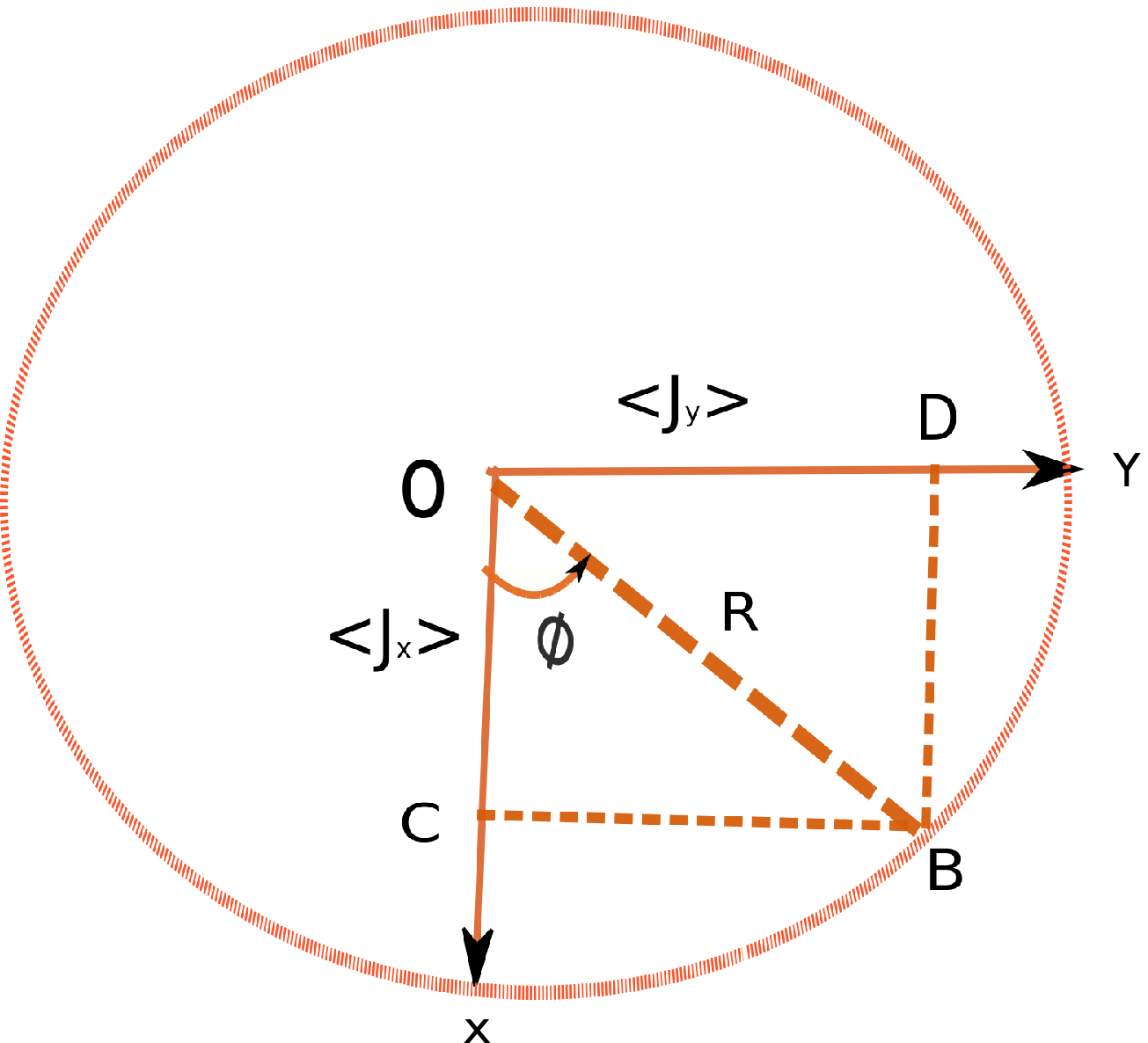} 
    \caption{Geometry in xy plane}\label{z1}
  \end{minipage}
    \begin{minipage}[b]{0.4\textwidth}
    \includegraphics[width=\textwidth]{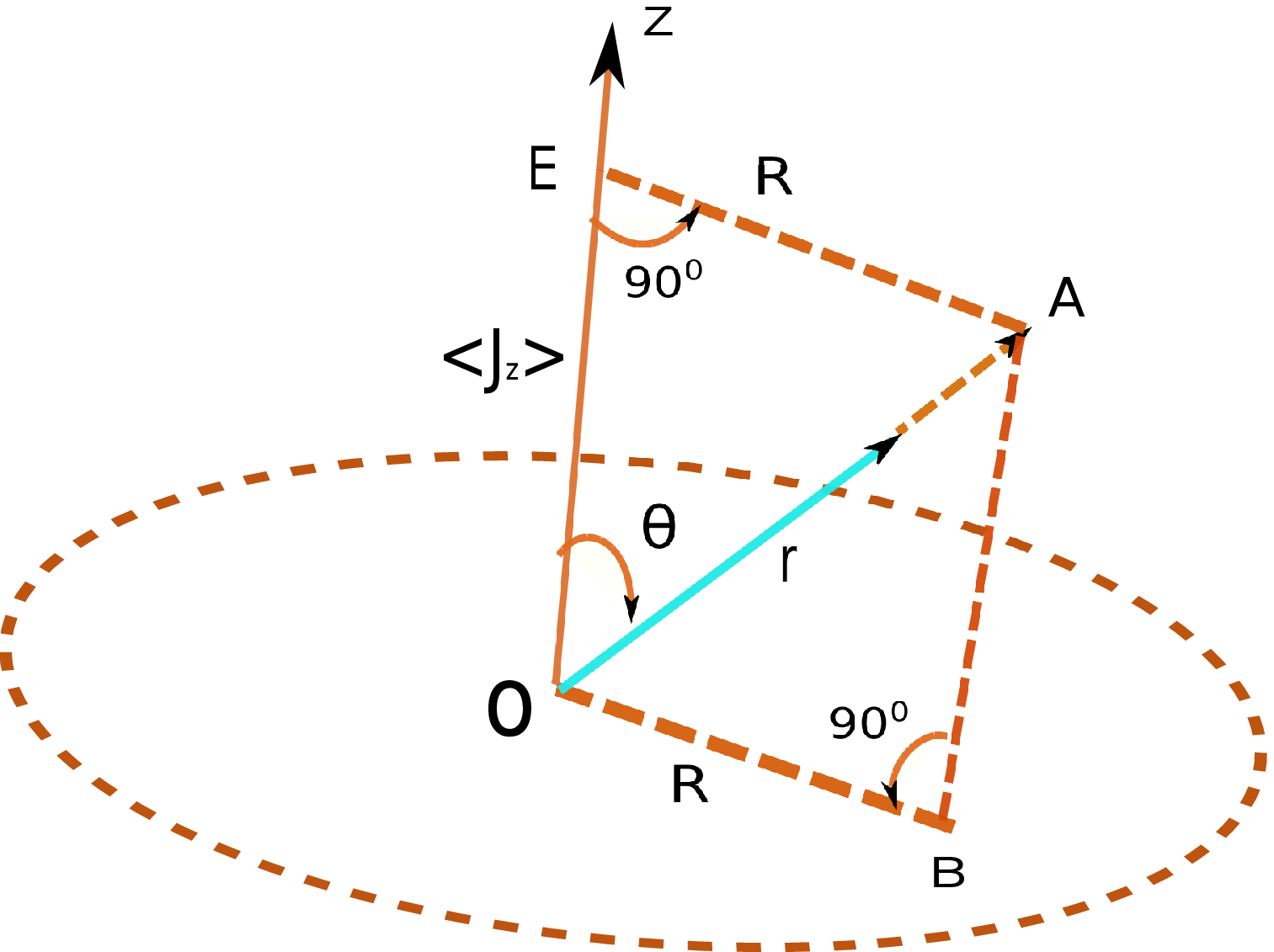}
    \caption{Geometry with mean wpin vector}\label{z2}
  \end{minipage}
\end{figure*}

Under this section we give the calculations of the variance $(\vartriangle J_{\varphi})^{2}$. To proceed, we define the variance as,
\begin{equation}
(\vartriangle J_{\varphi})^{2}=\langle J_{\varphi}^{2}\rangle-\langle J_{\varphi}\rangle^{2}. \label{var}
\end{equation}
By using the equation (\ref{e5}), we obtain the terms $\langle J_{\varphi}^{2}\rangle$ and $\langle J_{\varphi}\rangle^{2}$ as follows,
\begin{eqnarray}
J_{\varphi}^{2}=(J_{n_{1}}\cos\varphi+J_{n_{2}}\sin\varphi).(J_{n_{1}}\cos\varphi+J_{n_{2}}\sin\varphi)\\
=J_{n_{1}}^{2}\cos^{2}\varphi+ J_{n_{2}}^{2}\sin^{2}\varphi+\frac{1}{2} (J_{n_{1}}J_{n_{2}}+J_{n_{2}}J_{n_{1}})\sin 2\varphi \nonumber \\
=\frac{1}{2}(J_{n_{1}}^{2}-J_{n_{2}}^{2})\cos 2\varphi+ (J_{n_{1}}J_{n_{2}}
+J_{n_{2}}J_{n_{1}})\sin 2\varphi+ (J_{n_{1}}^{2}+J_{n_{2}}^{2})].
\end{eqnarray}
Taking the averages on both the sides we get,
\begin{eqnarray}
\langle J_{\varphi}^{2}\rangle=\frac{1}{2}[\langle J_{n_{1}}^{2}-J_{n_{2}}^{2}\rangle \cos 2\varphi+\langle J_{n_{1}}J_{n_{2}}
+J_{n_{2}}J_{n_{1}}\rangle \sin 2\varphi+\langle J_{n_{1}}^{2}+J_{n_{2}}^{2}\rangle]. \label{jp1}
\end{eqnarray}
Here we assume,
\begin{eqnarray}
M=\langle J_{n_{1}}^{2}-J_{n_{2}}^{2}\rangle\\
N=\langle J_{n_{1}}J_{n_{2}}
+J_{n_{2}}J_{n_{1}}\rangle \\
O=\langle J_{n_{1}}^{2}+J_{n_{2}}^{2}\rangle.
\end{eqnarray}
Hence the equation (\ref{jp1}) can be re written as,
\begin{equation}
\langle J_{\varphi}^{2}\rangle=\frac{1}{2}[M\cos2\varphi+N\sin\varphi+O].\label{v1}
\end{equation} 
Now focusing on the factor $\langle J_{\varphi}\rangle$, by using the equation (\ref{e5}) we get,
\begin{eqnarray}
\langle J_{\varphi}\rangle=\langle J_{n_{1}}\rangle \cos\varphi+\langle J_{n_{2}}\rangle \sin\varphi. \label{vjp}
\end{eqnarray}
Putting the values of the factors $\langle J_{n_{1}}\rangle$ and $\langle J_{n_{2}}\rangle$ from the equations (\ref{e2}) and (\ref{e3}) we further obtain,
\begin{dmath}
\langle J_{\varphi}\rangle=(-\langle J_{x}\rangle \sin\phi+\langle J_{y}\rangle\cos\phi)\cos\varphi+(\langle -J_{x}\rangle \cos \theta \cos \phi-\langle J_{y}\rangle \cos \theta \sin \phi+
\langle J_{z}\rangle \sin \theta)\sin\varphi. \label{ap}
\end{dmath}
Here we use geometric description to find out the values of the  factors $\langle J_{x}\rangle,\langle J_{y}\rangle$ and $\langle J_{z}\rangle$ used in the above equation. These are the components of mean vector along the x, y and z axis respectively. We refer the figure \ref{mf} and redraw two sub-figures \ref{z1} and \ref{z2}. Giving the look to the geometry in these figures, from the figure \ref{z1}, we find
\begin{eqnarray}
\cos\phi=\frac{\langle J_{x}\rangle}{R}\Rightarrow \langle J_{x}\rangle=R\cos\phi. \label{x} \\
\sin\phi=\frac{\langle J_{y}\rangle}{R}\Rightarrow 
\langle J_{y}\rangle=R\sin\phi \label{y}  \\
\tan\phi=\frac{\langle J_{y}\rangle}{\langle J_{x}\rangle} \label{tt} \\
R=\sqrt{\langle J_{x}\rangle^{2}+\langle J_{y}\rangle^{2}}. \label{RR}
\end{eqnarray}
From figure \ref{z2} we get the following results

\begin{eqnarray}
\sin\theta=\frac{R}{r} \Rightarrow R=r\sin\theta \label{ss}\\
\cos\theta=\frac{\langle J_{z}\rangle}{r}\Rightarrow \langle J_{z}\rangle=r\cos\theta  \label{cc} \\
r=\sqrt{\langle J_{x}\rangle^{2}+\langle J_{y}\rangle^{2}+\langle J_{z}\rangle^{2}}\neq R. \label{rr}
\end{eqnarray}
Where $r$ is the length of mean vector $\vec{J}_{mean}$, which is represented in the figures as a ray OA. The above expression obtained from the geometry shown in both the figures are independent from the state of the system and true for any state. We plug-in the values of the factors obtained in the equations (\ref{x},\ref{y},\ref{tt},\ref{RR},\ref{ss}),(\ref{cc}) and \ref{rr}, in equation (\ref{ap}), we get
\begin{equation}
(\langle J_\varphi \rangle=0) \Rightarrow (\langle J_\varphi \rangle^{2}=0) \label{jjp}
\end{equation} 
This is very interesting result and beauty of this result is that, it is true for any state of the system.
We plug-in the values from equations (\ref{v1}) and \ref{jjp} in equation (\ref{var}), we obtain the variance of the vector $\vec{J}_{\varphi}$ as below,
\begin{equation}
(\vartriangle J_{\varphi}^{2})=\frac{1}{2}[M\cos2\varphi+N\sin2\varphi+O]\label{fvar}
\end{equation}
The variance $(\vartriangle J_{\varphi}^{2})$ is the function of angle $\varphi$, we find the minimum and maximum value of the function over the angle $\varphi$, so doing first derivative of the function $(\vartriangle J_{\varphi}^{2})$ w.r.t the angle $\varphi$, we get,
\begin{eqnarray}
\frac{d}{d\varphi}=\frac{1}{2}[0+\frac{M}{2}(-\sin2\varphi)+\frac{N}{2}(\cos2\varphi)].
\end{eqnarray}
For maximization and minimization we put $\frac{d}{d\phi}=0$, which leads.
\begin{equation}
\tan2\varphi=\frac{N}{M}. \label{t2}
\end{equation}
The equation (\ref{t2}) further leads the conclusion as,
\begin{eqnarray}
\sin2\varphi=\pm\frac{N}{\sqrt{M^{2}+N^{2}}}, \quad
\cos2\varphi=\pm\frac{M}{\sqrt{M^{2}+N^{2}}}.\label{sc}
\end{eqnarray}
By putting the values from the equation (\ref{sc}) in equation.(\ref{fvar}), we obtain.
\begin{eqnarray}
(\vartriangle J_{\varphi})^{2}_{\pm}=\frac{1}{2}[O\pm \frac{M^{2}}{\sqrt{M^{2}+N^{2}}}+\frac{N^{2}}{\sqrt{M^{2}+N^{2}}}]\\
=\frac{1}{2}[O\pm \sqrt{M^{2}+N^{2}}].
\end{eqnarray}
As per the definition of spin squeezing we consider the minimum value of the variance along the vector $n_{\varphi}$, So the final expression for the variance of $J_{\varphi}$ is obtained as,
\begin{equation}
(\vartriangle J_{\varphi})^{2}_{-}=\frac{1}{2}[O- \sqrt{M^{2}+N^{2}}].
\end{equation}

\end{document}